# Synchronization and Control of Chaotic Spur Gear System Using Type-II Fuzzy Controller Optimized via Whale Optimization Algorithm

Farnaz Rezay Morsagh
Department of Electrical Engineering
Islamic Azad University, Science and Research Branch
Tehran, Iran
farnaz.rezaee1994@gmail.com

Mehdi Siahi
Department of Electrical Engineering
Islamic Azad University, Science and Research Branch
Tehran, Iran
mehdi.siahi@srbiau.ac.ir

Soudabeh Soleymani
Department of Electrical Engineering
Islamic Azad University, Science and Research Branch
Tehran, Iran
s.soleymani@srbiau.ac.ir

*Abstract*— Interval type-II Fuzzy Inference System (FIS) assumes a crucial role in determining the coefficients of the PID controller, thereby augmenting the controller's flexibility. Controlling chaotic systems presents inherent challenges and difficulties due to their sensitivity to initial conditions and the intricate dynamics that require precise and adaptive control strategies. This paper offers an exhaustive exploration into the coordination and regulation of a chaotic spur gear system, employing a Type-II Fuzzy Controller. The initial control parameters of the PID controller undergo optimization using the Whale Optimization Algorithm (WOA) to increase the overall system performance. The adaptability and strength of the suggested control system are tested in various scenarios, covering diverse reference inputs and uncertainties. The investigation comprehensively assesses the operational efficacy of the formulated controller, contrasting its performance with other methodologies. The outcomes highlight the impressive efficiency of the suggested strategy, confirming its supremacy in attaining synchronization and control within the turbulent spur gear system under demanding circumstances.

## I. Introduction

Chaotic systems have wide-ranging applications in disciplines such as engineering, robotics, computer science, and economics, emphasizing the importance of comprehending, controlling, and synchronizing their behavior. The intricate nature of chaotic systems becomes apparent due to their remarkable sensitivity to both initial conditions and nominal parameters. In simpler terms, minor adjustments to these conditions and parameters can exert a substantial influence on the trajectory and stability margins of chaotic systems, as highlighted in reference [1]. Consequently, tackling these challenges presents significant difficulty in successfully managing and stabilizing chaotic systems. This complexity underscores the critical need for in-depth understanding and advanced control strategies when dealing with chaotic systems in various practical applications [2].

Several conventional controllers, such as PID [2], adaptive [3], robust [4], and sliding mode [5]–[7], optimal [8] controller find application in the realm of controlling and synchronizing chaotic systems. These classical control techniques each offer distinct methodologies. PID controllers rely on proportional, integral, and derivative actions. The chaotic system is inherently characterized by nonlinearity and uncertainty, making it challenging to control, especially when subjected to various inputs. Intelligent controllers, such as fuzzy logic and neural network controllers, prove advantageous in handling the inherent nonlinearity of the system [9]. The amalgamation of a Fuzzy Inference System (FIS) with a traditional PID controller, as justified in [10], exhibits superior performance compared to a standalone conventional PID controller. Beyond their capability to manage nonlinear systems, these controllers offer additional benefits. These may include improved adaptability, robustness, and efficiency in responding to the unpredictable dynamics of chaotic systems, providing a more effective approach to control and synchronization challenges.

Several evolutionary optimization algorithms, including Gases Brownian Motion Optimization (GMBO) [11], the Imperialist Competitive Algorithm (ICA) [12], grasshopper optimization algorithm [13], Whale Optimization Algorithm (WOA) [14], and particle swarm optimization [15], have been applied for the optimization of controllers. These algorithms leverage principles inspired by natural phenomena or competitive strategies to fine-tune controller parameters. For instance, GMBO is inspired by Brownian motion, while the ICA mimics the dynamics of nations striving for dominance. Similarly, particle swarm optimization emulates the social behavior of particles. The utilization of such algorithms aims to enhance the efficiency and performance of controllers by exploring and optimizing the parameter space in innovative ways. These approaches exemplify the diverse and nature-inspired strategies employed to achieve optimal control in various systems.

This paper makes substantial contributions to the field of control strategies for chaotic spur gear systems. Firstly, by optimizing the initial control parameters of the Type-II Fuzzy Controller PID controller using the Whale Optimization Algorithm (WOA), the study enhances overall system performance. This optimization approach showcases a methodological advancement, indicating a commitment to refining control and synchronization strategies for chaotic systems. Secondly, the demonstrated adaptability and strength of the proposed control system, tested across diverse scenarios involving various reference inputs and uncertainties, reinforce its robustness. This adaptability is crucial for real-world applications where dynamic and unpredictable conditions are prevalent. Thirdly, the comprehensive assessment of operational efficacy through a thorough comparison with alternative methodologies demonstrates the paper's commitment to rigorously evaluating the proposed controller's performance. Lastly, the confirmed efficiency of the suggested strategy in achieving synchronization and control within a turbulent spur gear system, even under challenging circumstances, highlights the practical significance of the findings. Collectively, these contributions significantly advance the understanding and application of control strategies in the intricate domain of chaotic spur gear systems.

The following sections of this paper are organized as follows: Section 2 furnishes a detailed description of the system. Section 3 delineates the formulation of the proposed PID controller utilizing FIS optimized through WOA. To demonstrate the effectiveness of the proposed approach, Section 4 presents simulation results. Finally, Section 6 initiates a discussion and provides concluding remarks for the paper.

## II. SYSTEM DESCRIPTION

This section delves into the dynamic model of the chaotic spur gear system, as described by equation (1):

$$\dot{x}_1 = x_2$$
$$\dot{x}_2 = -2\varepsilon\mu x_2 + (0.1667x_1 - 0.1667x_1^3) + \varepsilon(f_m + f_e\Omega_e^2 \cos(\Omega_e\tau + \phi_e)) + u(t) \quad (1)$$

The system's state trajectory is notably sensitive to both nominal parameters and initial conditions. To analyze the phase plane and discern chaotic patterns within the system, the specified parameters in equation (1) are employed, with the following values: $f_m = 1, \varepsilon = 0.01, f_e = 30, \mu = 9, \Omega_e = 0.5$.

Figures 1-3 depict the system's phase portrait, showcasing the impact of initial conditions, where $x_1(0) = -2$, remains constant and $x_2(0) = 1$, and, $x_2(0) = 1.2, x_3(0) = 0.8$ are varied. This visualization aids in understanding the complex behavior and sensitivity of the chaotic spur gear system, highlighting the intricate dynamics influenced by both parameters and initial states.

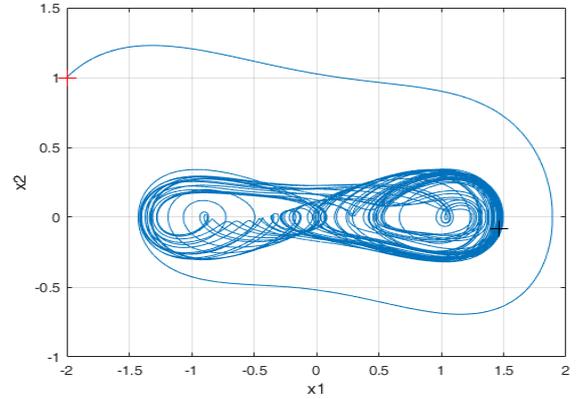

Fig. 1. The phase portrait of the spur gear system with the initial conditions $x_1(0) = -2, x_2(0) = 1$

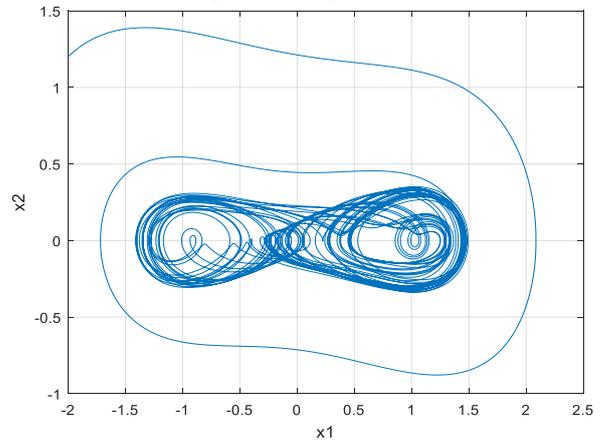

Fig. 2. The phase portrait of the spur gear system with the initial conditions $x_1(0) = -2, x_2(0) = 1.2$

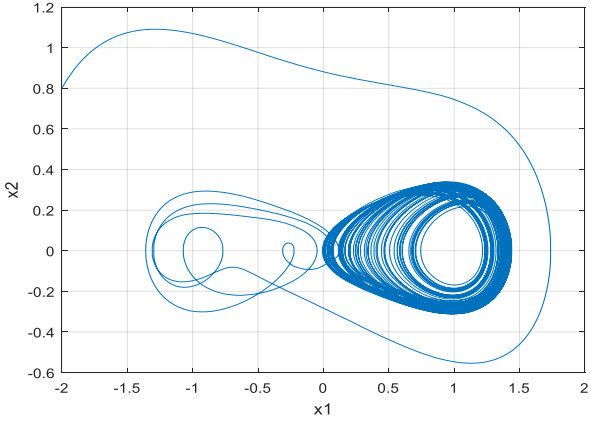

Fig. 3. The phase portrait of the spur gear system with the initial conditions $x_1(0) = -2, x_2(0) = 0.8$

As depicted in figures 1-3, a marginal alteration in the initial conditions of the chaotic spur gear system leads to a significant and unpredictable shift in its state trajectory as time progresses. This observation underscores the system's sensitivity to initial conditions, where even slight deviations can cause considerable variations in its dynamic behavior. The chaotic nature of the system amplifies this sensitivity, emphasizing the complexity and intricacy involved in predicting its trajectory over time. This inherent sensitivity to initial conditions necessitates precise control strategies to mitigate the unpredictable evolution of the system, highlighting the challenges associated with managing chaotic dynamics in practical applications. The nominal values and system details under examination in this study are outlined and provided in [16].

### III. THEORETICAL BASIS

#### A. Type-II Fuzzy Inference System

An interval type-II FIS is characterized by its use of interval type-II fuzzy sets. In contrast to type-I fuzzy sets where the membership function (MF) grade is crisp, in interval type-II fuzzy sets, the grade of MFs is fuzzy and defined as an interval. This unique attribute gives interval type-II fuzzy controllers an advantage in controlling systems without a precise mathematical model and in the presence of uncertainty, surpassing the capabilities of type-I fuzzy controllers [17]. In the context of interval type-II fuzzy sets, each fuzzy input set is transformed into upper and lower sets. The input of the interval type-II FIS, specifically the antecedent part, is then represented as. The utilization of interval type-II fuzzy sets in a FIS provides a more flexible and robust approach compared to traditional type-I fuzzy sets. The fuzzy nature of the MF grades, expressed as intervals, enables the system to handle imprecision and uncertainty more effectively [18]. This is particularly advantageous when dealing with control systems that lack precise mathematical models or operate in uncertain environments. In interval type-II fuzzy sets, the conversion of fuzzy input sets into upper and lower sets enhances the system's ability to interpret and respond to complex and uncertain information, making it a powerful tool for control in dynamic and unpredictable scenarios [19]. In interval type-II fuzzy sets, each fuzzy input set is converted to the upper and lower sets. The input of the interval type-II FIS (Antecedent part) is represented in (2-3).

$$\overline{\mu}(x) = \exp\left(-\frac{(x - c_k)^2}{\overline{\sigma}_k^2}\right) \quad (2)$$

$$\underline{\mu}(x) = \exp\left(-\frac{(x - c_k)^2}{\underline{\sigma}_k^2}\right) \quad (3)$$

where $x$ represents the input of the FIS, $k$ denotes the number of MFs, $c_k$, $\underline{\sigma}_k$ and $\overline{\sigma}_k$ stand for the center, lower, and upper parameters of the MFs of FIS, respectively. Additionally, $\overline{\mu}(x)$ and $\underline{\mu}(x)$ signify the MF grades. The corresponding outputs of the FIS, specifically the consequent part, are determined using the Biglarbegian method [20]. This method outlines a systematic approach to compute the FIS outputs based on the given input and MF parameters, ensuring a well-defined and coherent strategy for deriving the system's responses.

$$F_{T-II}(x|\theta) = m \frac{\sum_{l=1}^{M} \theta^l \overline{w}^l(x)}{\sum_{l=1}^{M} \overline{w}^l(x)} + (1 - m) \frac{\sum_{l=1}^{M} \theta^l \cdot \underline{w}^l(x)}{\sum_{l=1}^{M} \underline{w}^l(x)}$$

$, m \in [0\ 1]. \quad (4)$

where, $F_{T-II}(x|\theta)$ represents the interval type-II FIS outputs, with $\theta^l$ denoting the center of the consequent part. The parameter $M$ corresponds to the number of rules, and $\underline{w}^l(x)$ and $\overline{w}^l(x)$ represent, respectively, the products of lower and upper MF grades of the interval type-II FIS. The normalized vector of these products of MFs grades is detailed in equation (16). This mathematical representation outlines the computation of interval type-II FIS outputs and the associated parameters involved in the process, providing a clear framework for understanding the system's response based on the given inputs and rule conditions.

$$\zeta_U(x) = \frac{\overline{w}^l(x)}{\sum_{l=1}^{M} \overline{w}^l(x)} \quad , \quad \zeta_L(x) = \frac{\underline{w}^l(x)}{\sum_{l=1}^{M} \underline{w}^l(x)}. \quad (5)$$

In this context, $\zeta_U(x)$ and $\zeta_L(x)$ represent the normalized vectors of the upper and lower products of MFs grades, respectively. The parameter $m$ serves as the design parameter, indicating the effect ratio of the lower and upper MFs, and is a tunable parameter. Notably, as expressed in equations (4-5), the consequent MFs are of the singleton type. This conversion is integral to the application of the designed FIS and aids in adapting the system's responses based on the specified design parameters and input conditions. Through the application of equation (5), equation (7) undergoes a conversion process, facilitating a transformation in the representation of the system dynamics.

$$F_{T-II}(x|\theta) = m\,\theta^T\zeta_U(x) + (1-m)\theta^T\zeta_L(x)$$
$$= \theta^T\bigl(m\,\zeta_U(x) + (1-m)\zeta_L(x)\bigr). \quad (6)$$

$$\zeta_{T-II}(x) = m\,\zeta_U(x) + (1-m)\zeta_L(x). \quad (7)$$

In this context, $\zeta_{T-II}(x)$ denotes the normalized vector of products in the antecedent parts within the interval type-II FIS. Through a simplification process applied to equation (19), equation (20) is obtained. This simplification involves streamlining the mathematical expressions and relationships within the system, facilitating a more concise and interpretable representation. Equation (20) serves as a refined form, capturing the essential dynamics and interactions present in the interval type-II FIS, and thereby contributing to a more accessible understanding of the system's behavior.

$$F_{T-II}(x|\theta) = \theta^T \zeta_{T-II}(x). \quad (8)$$

The interval type-II FIS proposed in this study incorporates two inputs, namely the error and the derivative of the error, and produces five outputs representing the control parameters of the controller. These outputs correspond to the three coefficients associated with the proportional, integral, and derivative terms of the PID controller. The denotation of these outputs is as follows:

$$K_P = F_{T-II}\bigl(e,\dot{e}|\theta_{K_P}\bigr) = \theta_{K_P}^T \zeta_{T-II}(e,\dot{e}) \quad (9)$$
$$K_I = F_{T-II}\bigl(e,\dot{e}|\theta_{K_I}\bigr) = \theta_{K_I}^T \zeta_{T-II}(e,\dot{e}) \quad (10)$$
$$K_D = F_{T-II}\bigl(e,\dot{e}|\theta_{K_D}\bigr) = \theta_{K_D}^T \zeta_{T-II}(e,\dot{e}) \quad (11)$$

In the given context, $F_{T-II}(e,\dot{e}|\theta_{K_P})$, $F_{T-II}(e,\dot{e}|\theta_{K_I})$, and $F_{T-II}(e,\dot{e}|\theta_{K_D})$ represent the outputs of the FIS corresponding to the parameters of the proposed FPID (Fuzzy PID) controller. The parameters $\theta_{K_P}, \theta_{K_I}, \theta_{K_D}$ pertain to the consequent part of each output, while $\zeta_{T-II}(e,\dot{e})$ encompasses the antecedent parts of the FIS. These components collectively define the relationships between the inputs (error and derivative of error) and the control parameters of the proposed controller within the framework of the interval type-II FIS. According to (8-11), the signal control of the proposed controller is derived as

$$u_{FPID}(t) = F_{T-II}\bigl(e,\dot{e}|\theta_{K_P}\bigr)e(t) + \int F_{T-II}\bigl(e,\dot{e}|\theta_{K_I}\bigr)e(t)\,dt + F_{T-II}\bigl(e,\dot{e}|\theta_{K_D}\bigr)\dot{e}, \quad (12)$$

In this expression, $u_{FPID}(t)$ represents the signal control output of the fuzzy PID controller. The interval type-II FIS in use features nine singleton MFs within its consequent part. The parameter m is a tunable parameter crucial to the system, and its value must fall within the range of zero to one. The significance of m lies in its role in influencing the behavior of the system, allowing for fine-tuning and adaptation of the controller's response based on the specific requirements and characteristics of the controlled system [21] [22].

### B. Whale Optimization Algorithm

Evolutionary algorithms are widely recognized heuristic techniques employed for discovering optimal solutions to objective functions. The Whale Optimization Algorithm (WOA) is a novel addition to these algorithms, drawing inspiration from the hunting behavior of humpback whales [23]. Following the paradigm of population-based algorithms, WOA initiates with a set of random solutions, dividing the optimization process into two distinct phases: exploration and exploitation. During the exploration phase, the algorithm mimics the hunting behavior of whales by traversing in a spiral path around potential prey. In the exploitation phase, it employs a bubble-net feeding mechanism to capture and hunt down the prey. Each iteration involves acquiring the best candidate, and the remaining search agents adjust their positions toward this optimal agent. This dual-phase approach emulates the intricate and effective hunting strategies observed in nature, contributing to the algorithm's capability to efficiently explore and exploit the solution space. [24]. The mathematical model of searching prey is given in (13-14).

$$D = |C.X_{rand} - X(t)| \quad (13)$$
$$X(t+1) = X_{rand} - A*D \quad (14)$$

where the term $X$ denotes the position vector, representing either the search agent or the position of a whale in the WOA. Additionally, $X_{rand}$ signifies the random position vector of the search agent. These coefficients play a crucial role in the algorithm's dynamics, influencing the movement and updating mechanisms of the search agents or whales within the optimization process. The utilization of variable coefficients adds a level of adaptability to the algorithm, allowing it to dynamically adjust its behavior during each iteration based on the evolving search space and encountered solutions. The coefficients $A$ and $C$ are variable coefficient vectors, and their values are determined by the expressions outlined in equations (15) and (16).

$$A = 2ar - a \quad (15)$$

$$C = 2r \tag{16}$$

where $a$ is the parameter that in each iteration decreases linearly from 2 to 0, and $r$ is the random vector between 1 and 0. The encircling prey obtained according to (17) and (18).

$$D = |CX^*(t) - X(t)| \tag{17}$$
$$X(t+1) = X^*(t) - AD \tag{18}$$

The vector $X^*$ represents the position vector of the best search agent, and $t$ denotes the current iteration. The bubble-net attacking behavior is characterized by two approaches, as outlined in equation (19).

$$X(t+1) = \begin{cases} |X^*(t) - X(t)|e^{bl}\cos(2\pi l) + X^*(t) & p < 0.5 \\ X^*(t) - A*D & p > 0.5 \end{cases} \tag{19}$$

In equation (19), $l$ is a random number within the range [-1, 1], and $b$ defines the shape of a logarithmic spiral. The first scenario described in (19) involves updating the position through a spiral mechanism, while the latter scenario pertains to a shrinking encircling mechanism. The decision between these two approaches is determined by the random number $p$: if $p$ is less than 0.5, the algorithm employs the shrinking encircling mechanism; otherwise, the spiral updating position is utilized. The algorithm initiates with random solutions, and in each iteration, the solution's position is updated by the search agent. If A≥1, the search for prey is modeled according to equations (15-16), otherwise, the encircling prey through the shrinking mechanism is expressed by equations (17-18) [25].

As highlighted in the preceding section, the optimization involves determining the parameters of the Gaussian MFs in the antecedent part, specifically the center and sigma of these Gaussian MFs. Additionally, the initial values of the consequent MFs $\theta$ and $m$ as the coefficient of the interval type-II fuzzy are part of the optimization. Furthermore, the coefficients and fractional orders of the PID controller are obtained through the WOA The optimization problem's cost function involves the Integral Absolute Error (IAE) and Integral Time Absolute Error (ITAE). The constraints in the Fuzzy Proportional Integral Derivative (FPID) controller, represented by $\theta$, are outlined as follows.

$$0 \leq K_P, K_I, K_D \leq 10$$

## IV. SIMULATION RESULTS

In this Section, the WOA is utilized to compute the coefficients of the Proportional-Integral-Derivative (PID) controller, aiming for control parameters that minimize the combination of IAE and ITAE. Simultaneously, the algorithm is employed to determine the optimal placement of fuzzy sets in both the input and output. The outcomes from the design of three controllers—PID, Fuzzy PID (FPID), and interval type-II FPID—are subjected to comparison. Therefore, we examine the performance of the three controllers for four different scenarios, including various reference inputs, in the presence of uncertainties and different initial conditions. The scenarios presented in the simulation are initially discussed for the case where the system is without uncertainty. The next condition is considered for when the system is subject to uncertainty. The control signals and the first state variable of the system are plotted in various scenarios, and their performance indices are also presented in subsequent tables.

### A. Fisrt Scenario

In this scenario, the initial conditions are zero without the uncertainty. The reference input is $x_1(0) = 0$, $x_2(0) = 0$. The three controllers are evaluated and the results are shown in the Figures 4-5.

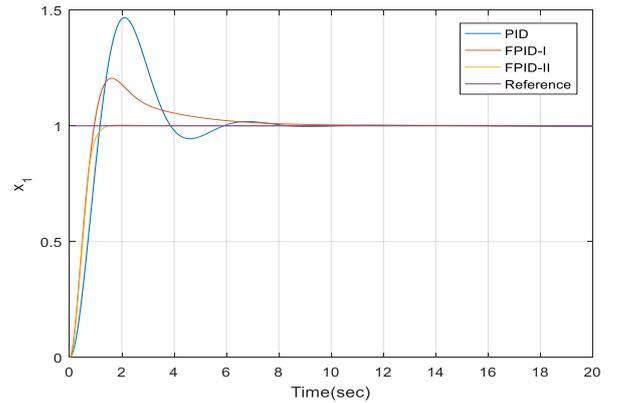

Figure 4: Comparison of the first state of the spur gear system using PID, FPID type-I, and FPID type-II for the first scenario.

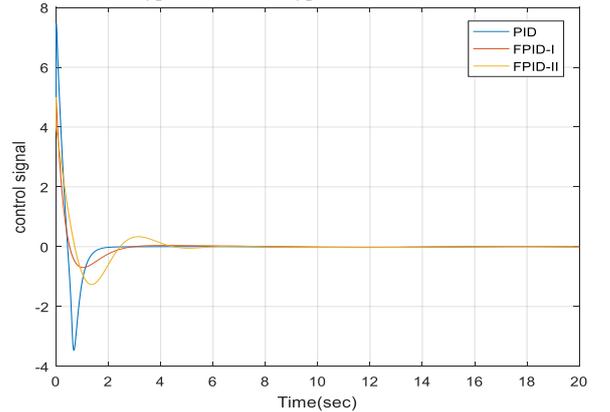

Figure 5: Comparison of the control signal using PID, FPID type-I, and FPID type-II for the first scenario.

### B. Second Scenario

The system's response for synchronizing with another chaotic system, where the master system has initial conditions of $x_1(0) = 1$, $x_2(0) = -1$, and the slave system is initialized with $x_1(0) = 0$, $x_2(0) = 0$. The presented results showcase the dynamic behavior and synchronization performance of the systems over time.

Table 1: IAE, ITAE for PID, FPID type-I, and FPID type-II for the first and second scenario

| Controller | IAE-First Scenario | ITAE-First Scenario | IAE-Second Scenario | ITAE-Second Scenario |
|---|---|---|---|---|
| PID | 1.4765 | 7.1869 | 2.8691 | 8.4465 |
| FPID type-I | 1.2401 | 5.6015 | 2.1234 | 6.5601 |
| FPID type-II | 1.0991 | 4.1271 | 1.5942 | 4.051 |

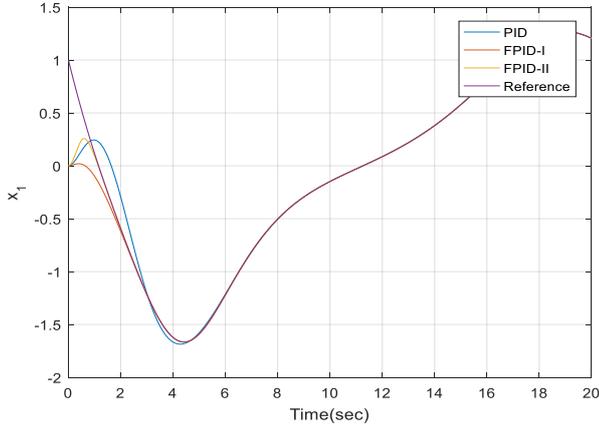

Figure 6: Comparison of the first state of the spur gear system using PID, FPID type-I, and FPID type-II for the second scenario.

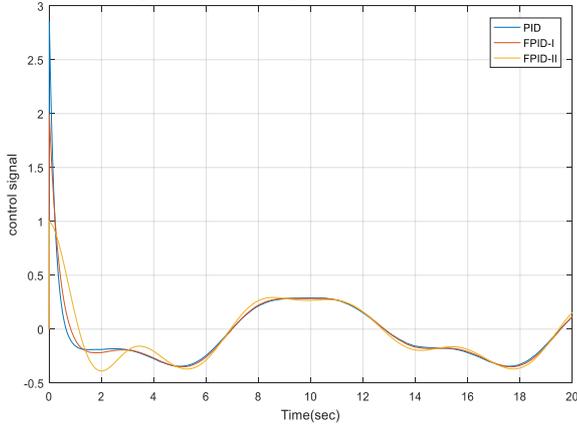

Figure 7: Comparison of the control signal using PID, FPID type-I, and FPID type-II for the second scenario.

### C. Third Scenario

In this particular setting, the system commences with initial conditions of zero, devoid of any uncertainty. The reference input is specified as $x_1(0) = -1$, $x_2(0) = 0.5$. The three controllers assessed under these conditions. The outcomes of this evaluation are visually presented in Figures 8 and 9.

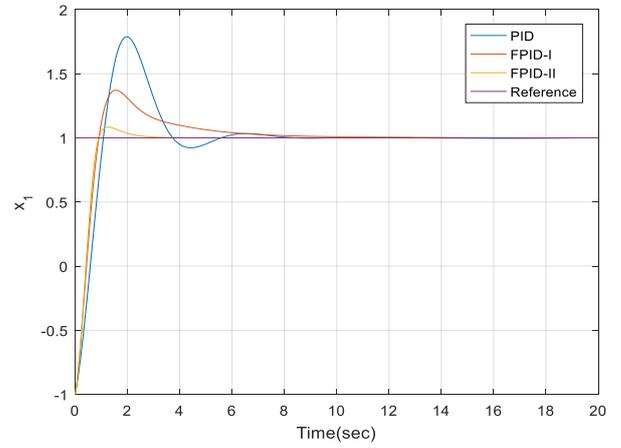

Figure 8: Comparison of the first state of the spur gear system using PID, FPID type-I, and FPID type-II for the third scenario.

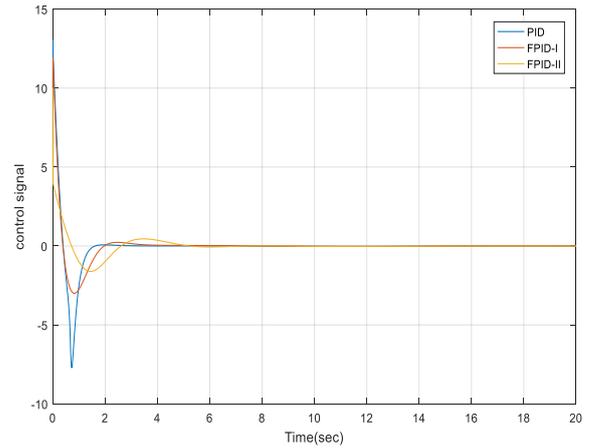

Figure 9: Comparison of the control signal using PID, FPID type-I, and FPID type-II for the third scenario.

### A. Fourth Scenario

The behavior of the system in synchronizing with another chaotic system is examined, where the master system starts with initial conditions $x_1(0) = 0$, $x_2(0) = 0$, and the slave system is initialized with $x_1(0) = -1$, $x_2(0) = 2$. The provided outcomes offer insights into the dynamic characteristics and synchronization effectiveness of these systems throughout the simulation.

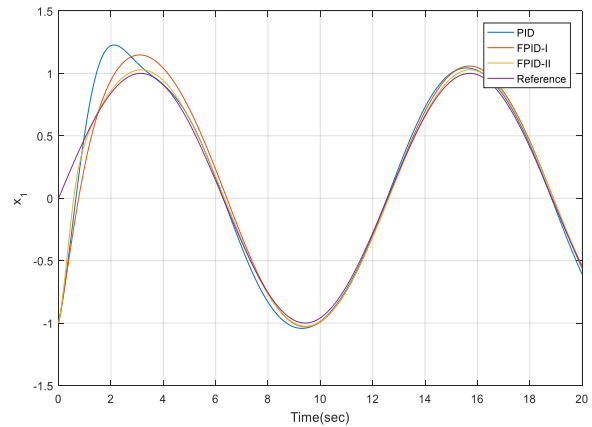

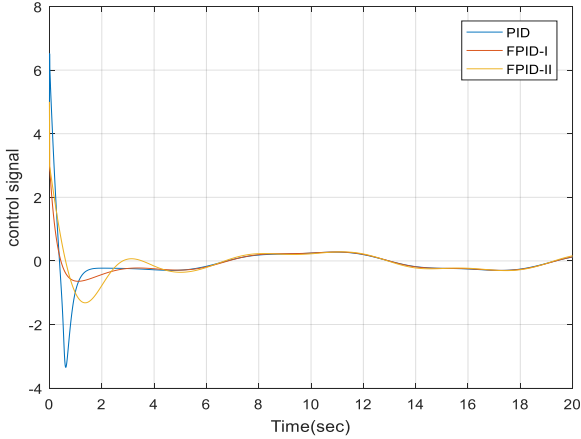

Figure 10: Comparison of the first state of the spur gear system using PID, FPID type-I, and FPID type-II for the fourth scenario.

Figure 11: Comparison of the control signal using PID, FPID type-I, and FPID type-II for the fourth scenario.

Table 2: IAE, ITAE for PID, FPID type-I, and FPID type-II for the third and fourth scenario

| Controller | IAE-Third Scenario | ITAE-Third Scenario | IAE-Fourth Scenario | ITAE-Fourth Scenario |
|---|---|---|---|---|
| PID | 2.9511 | 11.432 | 3.3513 | 11.3545 |
| FPID type-I | 2.6671 | 8.7351 | 2.6534 | 9.8831 |
| FPID type-II | 1.9431 | 5.3799 | 2.0942 | 5.5591 |

As depicted, the FPID controller demonstrates superior control and synchronization capabilities in comparison to the PID controller. This superiority stems from the dynamic adaptation of the FPID controller coefficients facilitated by the fuzzy system, allowing continuous adjustments to minimize errors at each moment. In contrast, PID controller coefficients remain constant, limiting their adaptability to changing system conditions. Furthermore, the second type of fuzzy system exhibits more effective performance in error reduction. The evaluation based on performance indices, such as ITAE and IAE, confirms that the FPID controller of the second type outperforms the other two controllers in terms of overall system performance. The dynamic and adaptive nature of the FPID controller, coupled with the enhanced error reduction capabilities of the second fuzzy system, contribute to its superior performance in controlling and synchronizing the chaotic system.

## V. CONCLUSION

In conclusion, this paper has explored the intricate dynamics of a chaotic spur gear system and proposed a control strategy utilizing a Type-II Fuzzy Controller. The Interval type-II Fuzzy Inference System (FIS) has been instrumental in enhancing the flexibility of the PID controller, addressing the inherent challenges associated with controlling chaotic systems. The optimization of PID controller parameters using the Whale Optimization Algorithm (WOA) has contributed to an overall improvement in system performance. The extensive evaluation of the suggested control system across various scenarios, encompassing diverse reference inputs and uncertainties, underscores its adaptability and robustness. Comparative analyses with alternative methodologies emphasize the effectiveness of the proposed approach, particularly in achieving synchronization and control within the turbulent spur gear system under demanding conditions. Furthermore, the study has highlighted the limitations of the traditional PID controller in chaotic systems, emphasizing the significance of dynamic and adaptive strategies. The introduction of the FPID controller, particularly the second type incorporating a fuzzy system, has proven to be a superior alternative. The dynamic adjustment of FPID coefficients through fuzzy logic enables continuous adaptation, addressing the evolving nature of chaotic systems. Evaluation based on performance indices, such as ITAE and IAE, confirms the superior performance of the FPID controller of the second type compared to both PID and the first type of FPID. This conclusively demonstrates the effectiveness of the proposed approach in achieving precise control and synchronization in chaotic systems, marking a significant advancement in the field of control theory.

In future endeavors, the exploration of advanced control strategies for the chaotic spur gear system holds promising avenues for further improvement. The incorporation of a Type-III FIS presents an intriguing prospect, offering enhanced modeling capabilities and adaptability compared to its predecessors. Type-III FIS, with its higher complexity and nuanced membership functions, may provide a more robust solution for the intricate dynamics of chaotic systems.